\documentclass[12pt,twoside,a4paper,amsmath,amssymb,showkeys,aps]{revtex4}

\usepackage[cp1251]{inputenc}
\usepackage[russian,english]{babel}

\usepackage[T2A]{fontenc}
\usepackage[pdftex]{graphics}
\usepackage{multirow}

\usepackage{graphics} 
\usepackage{epsfig}
\usepackage{amsfonts,amssymb,latexsym}
\usepackage{amsmath}
\usepackage{amsthm}
\usepackage{amstext}
\usepackage{amsmath}
\usepackage{latexsym}

\begin{document}
\DeclareGraphicsExtensions{.jpg,.pdf,.mps,.png} 

\title{
Polarization in quasirelativistic graphene model with
topologically non-trivial charge carriers
}

\author{Halina Grushevskaya }

\affiliation{Department of Computer Simulations,
Faculty of Physics \\
Belarusian State University, Minsk, Belarus}

\author{George Krylov}
\affiliation{Department of Computer Simulations,
Faculty of Physics \\
Belarusian State University, Minsk, Belarus}

\begin{abstract}
 Within the earlier developed high-energy-$\vec k\cdot \vec p$-Hamiltonian
  approach to describe graphene-like
 materials, the
simulations of band structure, non-Abelian Zak phases and
 complex conductivity of graphene  have been performed. The
quasi-relativistic graphene  model with a number of flavors   (gauge fields) $N_F=3$
 in two approximations (with and without  a pseudo-Majorana mass term) has been utilized as a ground
for the simulations. It has been shown that a Zak-phases set for
the non-Abelian Majorana-like excitations (modes)  in graphene
 is the cyclic group $\mathbf{Z}_{12}$ and this group is
deformed into a smaller one $\mathbf{Z}_8$  at sufficiently high
momenta due to a deconfinement of the modes.  Simulations of
complex longitudinal low-frequency conductivity  have been
performed with focus on effects of spatial dispersion. The spatial
periodic polarization in the graphene models with the pseudo
Majorana charge carriers is offered.
\end{abstract}


\maketitle


\section{Introduction}

Revolutionary progress in low dimensional physics is stipulated
primarily by the discovery of graphene and related materials.
Graphene belongs to bipolar materials that are characterized by strong correlations
due to many-body interactions.
The experimental facts \cite{Elias-et-al2012}
 testify that the Fermi velocity $v_F$ in valleys $K(K')$ (Dirac points)
 of a monolayer-graphene Brillouin zone is renormalized in the process
 of the Coulomb electron-electron interactions, and because of the
 weak screening the suspended-graphene dielectric constant $\epsilon_G $  remains moderate:
$\epsilon_G\approx 2.2$ and 5 for the small and large charge
concentrations $n\sim 10^9$ and $10^{12}$~cm$^{-2}$, respectively.
The renormalized $v_F$
grows with the growth of $ n $ in these experiments.
The experimental value of the effective dielectric constant $\epsilon_G$ for  bulk graphene
deposited on a hexagonal-boron-nitride support increases up to
$\approx 5$ due to dielectric-polarization support effects \cite{Zhao-Wyrick2015Science}.
Unconventional graphene superconductivity of non-phononic origin and another correlated
 insulator graphene states emerge in a twisted bilayer graphene (TBG) at some
``magic'' angles of rotation
of the graphene planes relative each other
due to the strong electron-electron interactions in the graphene also
\cite{Cao-et-al2018,Tao-et-al2021,Cao-et-al2021,Choi-et-al2021}.
Today, a filling-dependent band flattening caused by the strong interactions between electrons
in the bilayer graphene has be detected \cite{Choi-et-al2021}.
The fact that this phenomenon also occurs at the rotation angles, well above
the superconductivity magic angle, indicates the occurrence of the dielectric-polarization
process both in weak and strong screening regimes.
The screening of the Coulomb electron-electron interactions calculated using a
massless pseudo-Dirac fermion Hamiltonian within the Hartree-Fock approximation
favors the graphene superconductivity \cite{Guinea-Walet2018,Cea-et-al2019}.

To predict physically feasible results on 
the filling-dependent deformation of the  band structure at simulating one chooses 
unrealistically large screening, for example, $\epsilon _G\sim 66$ for the Hartree potential
\cite{Cea-et-al2019},
for  both the Hartree and Fock potentials, $\epsilon _G\sim 15$
\cite{Lewandowski-et-al2021}
 for the Hartree-Fock potential with a phonon-mediated pairing.
Thus, despite the fact that the Hartree and Fock potentials compete with each other,
the dielectric polarization in graphene is still predicted to be unwarranted high.

It can be assumed that collective particle-hole excitations are responsible
for the reduction of the dielectric polarization effects.
Then an excitonic insulating transition in the monolayer graphene would lead to
a significant increase in the value of the $ \epsilon_G $.
A quantum field theory of the graphene in an Eliashberg formalism predicts the excitonic
insulating transition \cite{Wang-et-al2011}.
However, the excitonic graphene gap is experimentally not observed \cite{Elias-et-al2012}.
Keldysh-type exciton states that can occur in electrically confined p-n (n-p) graphene
junctions could be responsible for experimentally-observed quasi-zero-energy states
of such a graphene quantum dot (GQD) at $\epsilon_G \sim 2.5$
\cite{Li-et-al2017}.
But, an estimate of the dielectric constant GQD which has been carried out by fitting
experimental local density of states by a massless pseudo Dirac--Weyl fermion
graphene model gives the value $ \epsilon_G \sim 1 $ and $ 2 $ for the ground and
other levels, respectively, as for the suspended graphene
\cite{Grush-PRB2021}.
It means that the exciton binding energy is very small to be observed.

The electron is a complex fermion, so if one decomposes it into
its real and imaginary parts, which would be Majorana fermions,
they are rapidly re-mixed by electromagnetic interactions.
However, such a decomposition could be reasonable for graphene
because of the effective electrostatic screening.
Pseudo Majorana  graphene fermion models become relevant also in connection with the discovery of the
unconventional superconductivity.
The pseudo Majorana fermions are topological vortical defects and
their statistics is non-Abelian one. A hindrance to describing
the vortex lattice lies in the impossibility to construct
maximally localized Wannier orbitals in a lattice site $i$ for a
band structure with topological defects owing to the presence of the defect in the site $i$.
Chiral superconductivity based on Majorana zero-energy edge modes is widely proposed as
a graphene superconductive model
(see \cite{Claassen-et-al2019,Tao-et-al2021a,Tao-et-al2021} and references therein).
But, according to the experiments performed in \cite{Choi-et-al2021,Xie-MacDonald2021,Cao-et-al2021}
the graphene superconductive states are nematic superconductive ones two-fold anisotropic
in the resistivity. Therefore, the feature of the nematic states
   is  broken gauge (spin/valley) and six-fold lattice rotational symmetries.
The violation of the chiral symmetry in the nematic-superconductivity phenomenon
casts doubt on the Majorana zero-energy edge modes in the graphene
(see \cite{Claassen-et-al2019,Tao-et-al2021a,Tao-et-al2021} and references therein).

The simplest massless pseudo Dirac fermion
Semenoff's model of the monolayer graphene \cite{Semenoff1984}
originated from a two-dimensional (2D) projection of the very old
non-relativistic graphite model proposed by Wallace in
\cite{Wallace} (see its further development in \cite{Neto2009} and reference therein).
Graphite and graphene are diamagnetic materials, and accordingly, one needs  the relativistic quantum
mechanics in order to correctly describe them.
One needs also a spin-orbit coupling (SOC) being a relativistic effect at describing Majorana fermions and
electron-hole pairs
as coupled Dirac-fermion states.  The spin-orbit coupling is
introduced into the non-relativistic graphene models in the form of
phenomenological corrections such as Rashba and Dresselhaus
spin-orbit couplings terms. But, a Dirac-mass Kane-Mele-term
\cite{Kane-Mele2005} originated from the non-zero SOC is
negligibly small one of the order of $10^{-3}$ meV for graphene at the
valleys $ K (K')$. In the case of the graphene monolayer without strain, the phenomenological
tight-binding model of the graphene superlattice with interlayer
interaction of the graphite type predicts the  flat bands \cite{Bistritzer2011},
 but unfortunately, parameters  of this  non-realistic model can not be adapted to experimental data.
The {\it ab initio} calculations predicted
 a gapped band structure of two-dimensional graphite
 though  the band structure is gapless one for three-dimensional (3D) graphite
  \cite{Grushevskaya-et-al1998}. But, in accordance with experimental data \cite{Elias-et-al2012}
though $v_F $ is diverged near $K$ no insulating phases emerge at
$E$ as low as 0.1~meV. Thus, the mass term for graphene can not be
of the Dirac type. Constructions of a mass term preserving chiral
symmetry for a graphene model of Majorana type are absent.
Thus, the massless pseudo Dirac fermion model is not applicable for a wide
range of phenomena in
 graphene physics such as the existence of topological currents in graphene superlattices
  \cite{Science346-2014Gorbachev}, signatures of Majorana excitation in graphene
\cite{PhysRevX5-2015San-Jose}, a sharp rise of Fermi velocity
value $v_F$ in touching valent  and conduction bands
\cite{Elias-et-al2012} and lack of excitonic instability
\cite{Wang-et-al2011}.

Moreover, the pure non-relativistic nature of the pseudo Dirac fermion model is in a
bad correspondence with modern {\it ab initio} software for band
structure simulations  like AbInit, FPLO, WIEN2k, VASP,
which is strongly related to 
quasirelativistic codes that proved to give results consistent
with experimentally observable properties of most materials \cite{Eschrig-et-al}.

So, the theoretical considerations based on the massless pseudo Dirac fermion graphene
model contradict the experimental studies that reveal the non-zero finite Fermi velocity
as well as bandwidths significantly larger than predicted.
 The dielectric polarization resulting in enhancement
of pairing and emerging correlated phenomena in the graphene remains elusive.
The problem of the graphene polarization is really challenge and is still unsolved one.

In \cite{myNPCS18-2015,NPCS18-2015GrushevskayaKrylovGaisyonokSerow,
Taylor2016,Grush-KrylSymmetry2016,our-symmetry2020}
we have captured the renormalization effect on the bandstructure
of the monolayer graphene at the level of a
quasirelativistic Dirac-Hartree-Fock approximation with $\epsilon_G=1$.
The  quasi-relativistic approximation in the relativistic quantum
mechanics  is known (see e.g., a review in
\cite{Kutzelnigg-et-al}) as the following procedure.
In terms of
the bispinor composed of two spinor components $\psi
=\left(\begin{array}{c}\varphi \\ \chi
\end{array} \right)$ the Dirac equation reads
$$\left( \hat{D}-m{{c}^{2}} \right)\psi =E\psi $$
where the operator $\hat D$ is written as
    $$\hat{D}=\left( \begin{array}{cc}
   m c^{2} +V & c \vec \sigma \cdot \vec p \\
   c \vec \sigma \cdot \vec p & - m c^2 + V
\end{array} \right)$$
with $\vec \sigma $ being the vector of the Pauli matrixes, $\vec
p$ is the momentum, $m$ is the electron mass, $c$ is the speed of
light, $V$ is some scalar potential.
To discard nonphysical solutions, the relationship  between upper
and lower bispinor components can be written in the form $\chi
=\hat{X}\varphi $, with $\hat X$ being a solution of the equation
$2m c^2\hat X=c\sigma p - [{\hat X},V] - c{\hat X}{\vec \sigma}
\cdot {\vec p}{\hat X}$.
One can omit two last terms in the right-hand side of the last
equation and gets $\hat X=\frac{\vec \sigma \cdot \vec p }{2mc}$.
In this case, the lower bispinor ($\chi $) component is of the
order of ${{c}^{-1}}$ of the upper ($\varphi $) one, that
corresponds to the use of the leading term in series expansion on
$c^{-1}$ for the original systems and is known as the
quasi-relativistic theory( or the limit).

We offer a quasirelativistic tight-binding Hamiltonian of massless
pseudo Majorana fermions in the monolayer graphene. The
pseudo Dirac fermion 2D model predicts the values of electrical and
magnetic characteristics that significantly differ from their experimental values,
and in this model there is no universal limit for the
low-frequency conductivity (``minimal'' direct-current (dc) conductivity) of graphene.
We  show that the polarization exists for the quasi-relativistic
model graphene with the pseudo Majorana modes.

Our goal in the paper is to study   the effects of topologically
non-trivial graphene Brillouin zone in dielectric properties of
the quasi-relativistic graphene model. Contributions to the
dielectric permittivity stipulated
by the presence of Majorana-like quasiparticle excitations will be calculated with
account of  spatial dispersion in the system.

\section{Methods}

\subsection{High-energy $\vec k\cdot \vec p$ Hamiltonian}
\begin{figure}[htbp]
\begin{center}
(a)\hspace{4cm} (b) 
\\
\includegraphics[width=3.5cm,height=3.5cm]{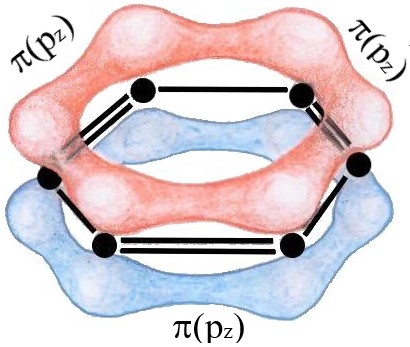}
\includegraphics[width=3.5cm,height=3.5cm]{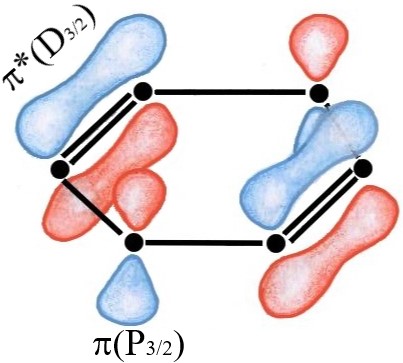}
\end{center}
\caption{  Graphene 
cell models with three dimers per hexagonal cell so that the
electron density of the dimers overlaps  (a) and  with two conjugated double  bonds  and one
``broken'' $\pi$-bond (b) that $\pi$- and
$\pi^*$-electrons are on terms
$P_{3/2}$ and $D_{3/2}$ with a total angular momentum
$J=3/2$. The
equality
of $P_{3/2}$- and
$D_{3/2}$-terms energies in the case %
of relativistic hydrogen-like atom provides stability of
configurations with two dimers.
}\label{color-models}
\end{figure}

Graphene is a 2D semimetal hexagonal carbon atomic layer, which is
comprised of two trigonal  sublattices $A,\ B$. Semi-metallicity
of graphene is provided by delocalization of
$\pi(\mbox{p}_z)$-electron orbitals  on a  hexagonal crystal cell
as is  shown
in Fig.~\ref{color-models}a.
 Since the energies of relativistic terms
 $\pi^* (D_{3/2}) $ and $\pi (P_{3/2}) $ of a hydrogen-like atom are equal to each other
 \cite{Fock}, there is an indirect exchange 
 through  
 $d$-electron states to break a dimer. Therefore,  a quasirelativistic model of
 monolayer graphene,
besides the configuration with three dimers per cell, also has a
configuration with two dimers and one broken  conjugate double
bond per the cell, as   is shown
in Fig.~\ref{color-models}b.
The basis set $\{\psi_{n_1},\psi_{n_2}\}$ for the $\pi$-, $\pi^*$-
orbitals is chosen in the following form:
\begin{equation}
\begin{split}
\psi_{\{n_1\}}= \psi_{\mbox{\small p}_z}(\vec r),\
\psi_{\{n_2\}}={1\over \sqrt{2}}\left(\psi_{\mbox{\small
p}_z}(\vec r\pm \vec \delta_i)+ \psi_{\mbox{\small p}_z}(\vec
r)\right),
\end{split}\label{basic-set}
\end{equation}
where $\psi_{\mbox{\small p}_z}$ is an atomic orbital of carbon
p$_z$-electron, $\vec \delta_i$, $i=1,2,3$ is a vector-radius of a
nearest neighbor for a carbon atom in the honeycomb lattice.

The high-energy $\vec k\cdot \vec p$ Hamiltonian of a
quasiparticle
 in the sublattice, for example, $A$
 reads \cite{our-symmetry2020}
\begin{eqnarray}
\left[\vec \sigma \cdot \vec p +  \vec \sigma \cdot \hbar
\left(\vec K_B - \vec K_A\right)
\right]\left| \psi^*_{BA}\right>  
-{i^2\over c}\Sigma_{AB}\Sigma_{BA}\widehat \psi^\dagger
_{-\sigma_A}\left|0,-\sigma \right> = E_{qu}
\widehat \psi^\dagger_{-\sigma_A} \left|0,-\sigma \right>, 
\label{Majorana-like-form} 
\end{eqnarray}
\begin{eqnarray}
 \left| \psi^*_{BA}\right> = \Sigma_{BA}\widehat \psi^\dagger _{-\sigma_A}\left|0,-\sigma \right>
\ \ \ \label{MwFun}
\end{eqnarray}
where  $\widehat \psi^\dagger _{-\sigma_A}\left|0,-\sigma \right>$
is a spinor  wave function (vector in the Hilbert space), $\vec
\sigma =\{\sigma_x, \sigma_y\}$ is the 2D vector of the Pauli
matrixes, $\vec p=\{p_x, p_y\}$ is the 2D momentum operator,
 $\Sigma_{AB}$, $\Sigma_{BA}$ are
relativistic exchange operators for sublattices $A, B$
respectively; $i^2\Sigma_{AB}\Sigma_{BA}$ is an unconventional
Majorana-like mass  term for a  quasiparticle in the sublattice
$A$, $\left| \psi^*_{BA}\right>$ is a spinor wave function of
quasiparticle in the sublattice $B$,  $\vec K_A(\vec K_B)$ denote
the graphene Dirac point  (valley) $\vec K$($\vec K'$) in the
Brillouin zone, $\hbar=h/(2\pi)$, $h$ is the Planck constant. A  small term $\hbar \vec \sigma\cdot\left( \vec K
_B - \vec K_A \right)\sim {\frac{h}{ a}}$ in
Eq.~(\ref{Majorana-like-form}) is a spin--valley-current coupling.
One can see that the term with the conventional Dirac mass in
(\ref{Majorana-like-form}) is absent. Since the exchange operators
transform a wave function from sublattice $A$ into $B$ and visa
versa in accord with Eq.\eqref{MwFun} the following expression
holds
\begin{equation} \nonumber
\begin{split}
\left|\tilde\psi^*_{BA}\right\rangle=\Sigma_{BA}\Sigma_{AB}\Sigma_{BA}
\left|\psi_A\right\rangle =\Sigma_{BA}^2\Sigma_{AB} \psi_A+
\Sigma_{BA}  [\Sigma_{AB},\Sigma_{BA}]\left|\psi_A\right\rangle\\
=
\Sigma_{BA}^2\{\Sigma_{AB} +
\Sigma_{BA}^{-1}[\Sigma_{AB},\Sigma_{BA}]\} \left|\psi_A
\right\rangle.
\end{split}
\end{equation}
Since the latter can be written in a form $\left| \tilde
\psi^*_{BA} \right\rangle= \Sigma_{BA}^2  \left|\tilde \psi_{A}
\right\rangle $
one gets the following property of the exchange operator matrix:
\begin{equation}
\left|\tilde \psi^*_{BA} \right\rangle \equiv \alpha ^{-1}
\Sigma_{BA} \left|\tilde\psi_{A} \right\rangle = \Sigma_{BA}^2
\left|\tilde\psi_{A}\right\rangle, \label{property}
\end{equation}
with some parameter $\alpha$. Due to the property
\eqref{property}, $\Sigma_{BA}
\left(\Sigma_{BA}\hat{\psi}^{\dagger}_{-\sigma_A}\right) = {\frac{1}{\alpha} }\Sigma_{BA} \hat{\psi}^{\dagger}_{-\sigma_A}$ and by  the
following notations
\begin{equation}
\vec \sigma_{AB}= \Sigma_{BA} \vec \sigma  \Sigma^{-1}_{BA},\ \vec
p_{BA}= \Sigma_{BA} \,\vec p\,  \Sigma^{-1}_{BA},\ \vec K^{BA}_{B}
-\vec K^{BA}_{A} = \Sigma_{BA} (\vec K_B  - \vec K_A
)\Sigma^{-1}_{BA},
\end{equation}
\begin{equation}
M_{BA}= i^2 \alpha \Sigma_{BA} \Sigma_{AB}, \ M_{AB}= i^2 \alpha
\Sigma_{AB} \Sigma_{BA}. \label{mass-operator}
\end{equation}
 Eq.~(\ref{MwFun}) can be rewritten  as
\begin{eqnarray}
\left[\vec \sigma_{AB}\cdot \left(\vec p_{BA} + \hbar (\vec
K^{BA}_B - \vec K^{BA}_A)\right) -{\frac{1}{ c}} M_{BA}\right]
\Sigma_{BA} \hat{\psi}^{\dagger}_{-\sigma_A}  \left | 0,
-\sigma\right> =\hat v_F^{-1}E_{qu}\Sigma_{BA}
\hat{\psi}^{\dagger}_{-\sigma_A}\left|0,-\sigma \right>
. 
\label{gapped-Majorana1}
\end{eqnarray}
The equation similar to (\ref{gapped-Majorana1}), can be also
written for the
sublattice $B$. As a result, one gets the
equations of motion for a Majorana bispinor
$(\left|\psi_{AB}\right \rangle, \left|\psi^*_{BA}\right
\rangle)^T$
 \cite{Grush-KrylSymmetry2016,myNPCS18-2015}:
\begin{eqnarray}
\left[\vec \sigma_{2D}^{BA}\cdot \vec p_{AB} -c^{-1} 
M_{AB} \right]\left| \psi_{AB}\right\rangle =
i {\frac{\partial}{\partial t}}\left| \psi^*_{BA}\right \rangle , 
\label{Majorana-bispinor01}  \\
\left[\vec \sigma_{2D}^{AB}\cdot \vec p\,^*_{BA}
-c^{-1}\left( M_{BA} 
\right)^*\right] \left|\psi_{BA}\right\rangle 
  = - i {\frac{\partial}{\partial t}}\left|
\psi^*_{AB}\right \rangle . 
\quad\label{Majorana-bispinor02}
\end{eqnarray}
Then, the exchange interaction term $\Sigma_{rel}^{x}$  is
determined as \cite{NPCS18-2015GrushevskayaKrylovGaisyonokSerow}
\begin{eqnarray}
\Sigma_{rel}^{x}\left(
\begin{array}{c}
\widehat {\chi } ^{\dagger}_{_{-\sigma_{_A}} }(\vec r) \\
\widehat {\chi }^\dagger _{\sigma_{_B}}(\vec r)
\end{array}
\right)\left|0,-\sigma \right> \left|0,\sigma \right> =
 \left(
\begin{array}{cc}
0&  \Sigma_{AB}
\\
\Sigma_{BA} & 0
\end{array}
\right)
\left(
\begin{array}{c}
\widehat {\chi }^{\dagger}_{-\sigma_{_A} } (\vec r) \\
\widehat {\chi} ^\dagger _{\sigma_{_B}}(\vec r)
\end{array}
\right)\left|0,-\sigma \right> \left|0,\sigma \right>
\label{exchange}
, \\ 
\Sigma_{AB} \widehat {\chi }^\dagger _{\sigma_{_B}}(\vec
r)\left|0,\sigma \right>
\nonumber \\
= \sum_{i=1}^{N_v\,N_c}\int { d \vec r_i} \widehat {\chi }^\dagger
_{\sigma_i{^B}}(\vec r)\left|0,\sigma \right> \Delta _{AB} \langle
0,-\sigma_i|{\widehat \chi}^\dag_{-\sigma_i^A} (\vec r_i) V(\vec
r_i -\vec r) {\widehat \chi}_{-\sigma_B}(\vec
r_i)|0,-\sigma_{i'}\rangle , \ \ \label{Sigma-AB}
\\ 
 \Sigma_{BA}
\widehat {\chi }^{\dagger}_{_{-\sigma_{_A}} } (\vec r)
\left|0,-\sigma \right>
\nonumber\\
=\sum_{i'=1}^{N_v\,N_c}\int { d \vec r_{i'}} \widehat {\chi
}^{\dagger}_{_{-\sigma_{i'}^A} } (\vec r) \left|0,-\sigma \right>
\Delta _{BA} \langle 0,\sigma_{i'}|{\widehat
\chi}^\dag_{\sigma_{i'}^B} (\vec r_{i'}) V(\vec r_{i'} -\vec r)
{\widehat \chi}_{_{\sigma_A}}(\vec r_{i'})|0,\sigma_i\rangle . \ \
\label{Sigma-BA}
\end{eqnarray}
Here $N_v$ and $N_c$ are a number of valent electrons in one primitive-lattice cell and a number of the cells;
interaction ($2\times 2$)-matrices $\Delta _{AB}$ and $\Delta_{BA}$ are gauge fields (or components of a gauge field).
In the case of the basic set (\ref{basic-set}),  vector-potentials for
these gauge fields are determined by the phases $\alpha_{ 0}$ and $\alpha_{\pm, k}$, $k=1,\ 2,\ 3$ of
 wave functions $\psi_{\mbox{\small p}_z}(\vec r)$ and $\psi_{\mbox{\small p}_z, \pm \vec
\delta_k}(\vec r)$,
$k=1,\ 2,\ 3$ respectively that
the exchange interaction $\Sigma_{rel}^{x}$ (\ref{exchange}) in
accounting of the nearest lattice neighbours  for a tight-binding
approximation
reads 
\cite{Taylor2016,myNPCS18-2015,NPCS18-2015GrushevskayaKrylovGaisyonokSerow}
\begin{eqnarray}
 \Sigma_{AB}
 ={\frac{1}{ \sqrt{2}(2\pi)^{3}}}
e^{-\imath (\theta_{k_{A}}-\theta_{K_B})}
\sum_{i=1}^{3} \exp\{\imath [\vec K^i_{A} - \vec q_i ] \cdot \vec
\delta_i\} \int  V(\vec r) d {\vec r}
\nonumber \\
 \times \left(
\begin{array}{cc}
\sqrt{2}\psi_{\mbox{\small p}_z} (\vec r ) \psi^*_{\mbox{\small
p}_z, - \vec \delta_i} (\vec r )
 &
\psi_{\mbox{\small p}_z} (\vec r ) [\psi^*_{\mbox{\small
p}_z}(\vec r)
+ \psi^*_{\mbox{\small p}_z, - \vec \delta_i}(\vec r)]\\
\psi^*_{\mbox{\small p}_z, - \vec \delta_i} (\vec r)
[\psi_{\mbox{\small p}_z, \vec \delta_i}(\vec r)+
\psi_{\mbox{\small p}_z}(\vec r)] & {[\psi_{\mbox{\small p}_z,
\vec \delta_i}(\vec r)+ \psi_{\mbox{\small p}_z}(\vec r)]
[\psi^*_{\mbox{\small p}_z}(\vec r) +\psi^*_{\mbox{\small p}_z, -
\vec \delta_i}(\vec r)] \over \sqrt{2}}
\end{array}
\right)
, \label{Sigma-AB3}
\\ 
\Sigma_{BA}
= {1\over \sqrt{2}(2\pi)^{3}} e^{-\imath
(\theta_{K_A}-\theta_{K_B})}
\sum_{i=1}^{3} \exp\{\imath [\vec K^i_{A} - \vec q_i ] \cdot \vec
\delta_i\}  \int  V(\vec r) d {\vec r}
\nonumber \\
 \times \left(
\begin{array}{cc}
{[\psi_{\mbox{\small p}_z, \vec \delta_i}(\vec r)+
\psi_{\mbox{\small p}_z}(\vec r)] [\psi^*_{\mbox{\small p}_z}(\vec
r) +\psi^*_{\mbox{\small p}_z, - \vec \delta_i}(\vec r)] \over
\sqrt{2}}
 &
- \psi^*_{\mbox{\small p}_z, - \vec \delta_i} (\vec r )
[\psi_{\mbox{\small p}_z, \vec \delta_i}(\vec r)+
\psi_{\mbox{\small p}_z}(\vec r)]
\\
- \psi_{\mbox{\small p}_z} (\vec r ) [\psi^*_{\mbox{\small
p}_z}(\vec r) + \psi^*_{\mbox{\small p}_z, - \vec \delta_i}(\vec
r)] & \sqrt{2}\psi_{\mbox{\small p}_z} (\vec r )
\psi^*_{\mbox{\small p}_z, - \vec \delta_i} (\vec r)
\end{array}
\right)
 \label{Sigma-BA3}
\end{eqnarray}
where the origin of the reference frame is located at a given site
on the sublattice $A$($B$),
 $V(\vec r)$ is the 3D Coulomb potential,
designations $\psi_{\mbox{\small p}_z,\ \pm \vec \delta_i}(\vec
r)$, $\psi_{\mbox{\small p}_z,\ \pm \vec \delta_i}(\vec
r_{2D})\equiv \psi_{\mbox{\small p}_z}(\vec r\pm \vec \delta_i)$,
$i=1,2,3$ refer to
atomic orbitals  of p$_z$-electrons with 3D radius-vectors $\vec
r\pm \vec \delta_i$ in the neighbor lattice sites
$\vec \delta_i$, nearest to the reference site
;
 $\vec r\pm \vec \delta_i$ is the p$_z$-electron 3D-radius-vector.
Elements of the matrices $\Sigma_{AB}$ and $\Sigma_{BA}$
include bilinear combinations of the wave functions so that
their phases  $\alpha_{ 0}$ and $\alpha_{\pm, k}$, $k=1,\ 2,\ 3$
enter into $\Delta_{AB}$ and $\Delta_{BA}$ from
(\ref{Sigma-AB} and \ref{Sigma-BA}) in the form
\begin{eqnarray}
\left|\psi_{\mbox{\small p}_z}\right| \left|\psi_{\mbox{\small
p}_z,\ \pm \vec \delta_k}\right|
 \exp\left\{\imath \left( \alpha_0 - \alpha_{\pm,
k}\right)\right\} 
\equiv \left|\psi_{\mbox{\small p}_z}\right|
\left|\psi_{\mbox{\small p}_z,\ \pm \vec \delta_k}\right| \Delta
_{\pm,k}
. \label{phase_variation}
\end{eqnarray}
Due to the fact that phases are included into (15) only with their differences,
an effective number $N_F$ of flavors in our gauge field theory is equal to 3,
separately for  holes  and electrons (signs plus and minus in (15) refer to hole and electrons, respectively).
Then owing to
translational symmetry we  determine the  gauge fields $\Delta _{\pm,i}$ in
Eq.~(\ref{phase_variation})
in the following form:
\begin{equation}
\label{c-alpha} \Delta _{\pm,i}(q) = \exp\left(\pm \imath c_\pm
(q)({\vec q}\cdot{\vec \delta_i})\right).
\end{equation}
Substituting the relative phases \eqref{c-alpha}
of particles
and holes
into \eqref{Sigma-AB3} one gets the exchange interaction operator
$\Sigma_{AB}$
\begin{eqnarray}
 \Sigma_{AB} ={1\over \sqrt{2}(2\pi)^{3}}
e^{-\imath (\theta_{k_{A}}-\theta_{K_B})}
 \left(
\begin{array}{cc}
\Sigma_{11} &\Sigma_{12}\\
\Sigma_{21}& \Sigma_{22}
\end{array}
\right)  \label{Sigma-AB3-second-approximation}
\end{eqnarray}
with the following matrix elements:
\begin{eqnarray}%
\Sigma_{11}= \sqrt{2} \left\{\sum_j I_{11}^j 
\Delta _{-,j}(q) \exp\{\imath [\vec K^j_{A} - \vec q ] \cdot \vec
\delta_j\} \right\},
\label{Sigma-AB11-second-approximation} \nonumber \\
\Sigma_{12}=   \left\{\sum_j \left( I_{12}^j +  I_{11}^j
\Delta _{-,j}(q)
 \right)
\exp\{\imath [\vec K^j_{A} - \vec q ] \cdot \vec \delta_j\}
\right\} , 
\nonumber \\
\Sigma_{21}=   \left\{\sum_j \left(I_{21}^j 
\Delta _{+,j}(q)\Delta _{-,j}(q)
+  I_{11}^j 
\Delta _{-,j}(q) \right) \exp\{\imath [\vec K^j_{A} - \vec q]
\cdot \vec \delta_j\}
\right\},  
\nonumber
\\
\Sigma_{22}= {1 \over \sqrt{2}}   \left\{\sum_j \left( I_{22}^j 
\Delta _{+,j}(q)
 +
I_{12}^j+ I_{21}^j 
\Delta _{+,j}(q)\Delta _{-,j}(q)
+I_{11} ^j 
\Delta _{-,j}(q) \right) \exp\{\imath [\vec K^j_{A} - \vec q ]
\cdot \vec \delta_j \}
\right\} 
\nonumber 
\end{eqnarray}
where $ I_{n_i m_k}^j = \int  V(\vec r)  \psi_{\mbox{\small
p}_z+n_i\vec \delta_j} {\psi^*}_{\mbox{\small p}_z-m_k\vec
\delta_j}  \ d\vec r
$, $i,k=1,2$; $(n_1,m_1)=(0,1)$, 
$(n_1,m_2)=(0,0)$, 
$(n_2,m_1)=(n_2,m_2)=(1,1)$. 
There are  similar formulas for $\Sigma_{BA}$.
%

Accordingly to (\ref{c-alpha})  eigenvalues  $E_i, \ i=1,\ldots, 4$
of  the $4\times 4$ Hamiltonian
(\ref{Majorana-bispinor01}, \ref{Majorana-bispinor02})   are functionals of $c_\pm
$.
To eliminate arbitrariness in the choice of the phase factors
$ c_ \pm $ 
one needs a gauge condition for the gauge fields. 
The eigenvalues $E_i, \ i=1,\ldots, 4$ are real because the system
of equations (\ref{Majorana-bispinor01},
\ref{Majorana-bispinor02}) can be transformed to
Klein--Gordon--Fock equation \cite{Grush-KrylSymmetry2016}.
Therefore we 
impose the gauge condition 
 as a  requirement on the absence of imaginary parts in the eigenvalues
$E_i, \ i=1,\ldots, 4$ of  the Hamiltonian
(\ref{Majorana-bispinor01}, \ref{Majorana-bispinor02}):
\begin{equation}\label{sys}
\Im m(E_{i})=0, \ i=1,\ldots , 4.
\end{equation}
To satisfy the condition (\ref{sys}) in the momentum space we  minimize  the  function $f(c_+
, c_-
)=\sum_{i=1}^4\left|\Im m \ E_{i}\right|$   absolute minimum of
which coincides with the solution of the system (\ref{sys}). The
bandstructures determined by  the sublattice Hamiltonians without the pseudo-Majorana mass term are
the same. Therefore when neglecting the pseudo-Majorana mass term, one can choose  the cost function
 $f={2} \sum_{i=1}^2\left|\Im m \ E_{i}\right|$. For the non-zero
mass case, we assume the same form of the function $f$ due to the
smallness of the mass correction.

\subsection{Non-Abelian-Zak-phase analysis of emerging polarization}
Topological defect pushes out a charge carrier from its location.
The operator of this non-zero displacement
presents a projected position operator ${\cal P}  {\vec r} {\cal
P}$ with the projection operator ${\cal P}= \sum_{n=1}^{N_B}
\left|\psi_{n,\vec k}\right\rangle \left\langle \psi_{n,\vec
k}\right|$ for the occupied subspace of states $\psi_{n,\vec
k}(\vec r)$. Here $N_B$ is a number of occupied bands, $\vec k$ is a
momentum.
Eigenvalues of ${\cal P}  {\vec r} {\cal P}$ are called 
Zak phase 
\cite{Zak89}.  The  Zak phase coincides with a phase
\begin{eqnarray}
\gamma_{mn}=i\int\limits_{C(\vec{k})}\left<\psi_{m,\vec{k}}\left|\nabla_{\vec{k}}\right|\psi_{n,\vec{k}}
\right>\cdot d\vec{k}, \quad n,m=1,\ldots, N_B \label{non-A-ZP}
\end{eqnarray}
of a Wilson loop ${\cal W}^{mn} =\mbox{T }\exp(i  \gamma_{mn}) $
being a path-ordered (T)  exponential with the integral over a
closed contour $C(\vec{k})$ \cite{PhysRevX6-2016Muechler}. We
 discretize the Wilson loop   by Wilson lines ${\cal W}^{mn}_{k_{i+1},k_i}$
\cite{our-symmetry2020}:
\begin{eqnarray}
{\cal W}^{mn}=\prod_{i=0}^{N_{\cal W}\to \infty}{\cal
W}^{mn}_{\vec k_{i+1},\vec k_i}=\prod_{i=0}^{N_{\cal W}\to \infty}
\left\langle \psi^*_{m,\vec k_{i+1}}\left| \right. \psi_{n,\vec
k_{i}}\right\rangle .
 \label{WilsonLoop}
\end{eqnarray}
Here momenta $\vec k_i$, $i=0,1,\ldots, N_{\cal W} $ form a
sequence of the points on a curve (ordered path), connecting initial and final points in the
Brillouin zone: $\vec k_i=\vec k_0+ \sum_{j=1}^{i}\Delta \vec
k_{j,j-1}$, $\Delta \vec k_{j,j-1}=\vec k_{j}-\vec k_{j-1}\to 0$
and $\vec k_{N_{\cal W}}=\vec k_0$; $\psi_{n,{\vec k_i}}$,
$n=1,\ldots,N_B$ are eigenstates of a model Hamiltonian.

We consider the parallel transport of filled Bloch waves around
momentum loops $\vec l$ because the basis of Wannier functions
generated only by the occupied Bloch eigenstates. Global
characterization of all Dirac touching is possible with a
non-Abelian Zak invariant defined over a non-contractible momentum
loop \cite{Zak89}. Therefore, instead of the closed contour, we take a
curve $C(\vec{k})$ being one side $\vec l (k_y)$ of the
equilateral triangle of variable size (defined by the value of
$k_y$ component of the wavevector $\vec k$) with the
coordinate-system origin in the Dirac $K(K')$-point.
The $N_B$ phases are defined then as arguments of the eigenvalues of the
Wilson loop. One chooses $N_{\cal W}$ ($N_{\cal W}=500$) that a
``noise'' in output data  is sufficiently small to observe the discrete values of Zak phases.
In our calculation  of (\ref{WilsonLoop}) for the Hamiltonian without the pseudo-Majorana mass term,
 a number $N_B$ of bands is equal to four ($N_B=4$):
 two electron and hole valent bands and two electron and hole conduction bands
 (see simulation results in Section~2.1).

\subsection {Non-Abelian currents in quasi-relativistic graphene model}

Conductivity can be considered as a coefficient linking the
current density with an applied electric field in a linear regime of
response. To reach the goal, several steps should be performed.
First, one has to subject the system to an electromagnetic field,
this can be implemented by standard change to canonical momentum $
\vec p \to \vec p - { e\over c}\vec A$ in the Hamiltonian where
$\vec A$ is a vector-potential of the field, $e$ is the  electron charge.
Then, one can find a quasi-relativistic current 
\cite{Davydov} 
of charge carriers 
in graphene  as:
\begin{eqnarray}
j_i^{SM} =c^{-1}j_i\equiv j_i^{O}+j_i^{Zb}+j_i^{so}, 
\nonumber \\
j_i=e \chi^{\dagger}_{+\sigma_{_B} } (x^+) v^i_{x^{+}x^-}
\chi_{+\sigma_{_B} } (x^-)
 -{e^2 A_i\over c\,
 M_{AB}}
 \chi_{+\sigma_{_B} }^\dagger (x^+) \chi_{+\sigma_{_B} }(x^-)
\nonumber \\
 + {e \hbar \over 2
 M_{AB}}
\left[
 \vec \nabla \times \chi_{+\sigma_{_B} }^\dagger (x^+)\vec \sigma
 \chi_{+\sigma_{_B} }(x^-)
\right]_i, \ i=1,2 . \label{graphene-quasirel-current}
\end{eqnarray}
Here
\begin{eqnarray}
x^{\pm}=x \pm\epsilon ,\ x=\{\vec r,\ t_0 \},\ \vec r=\{x,y\}, \
t_0=0,\ \epsilon \to 0 ;\label{current-limits}
\end{eqnarray}
$v^i_{x^{+}x^-} $ is the velocity operator determined by a
derivative of the Hamiltonian
(\ref{Majorana-bispinor01},\ref{Majorana-bispinor02}),
$\chi_{+\sigma_{_B} } (x^+)$ is the secondary quantized fermion
field,
the terms 
$j_i^{O},\ j_i^{Zb},\ j_i^{so}$, $i=x,y$  describe an ohmic
contribution which satisfies the Ohm law  and contributions of the
polarization and magneto-electric effects respectively.
A potential-energy operator $V$ for interaction between the
secondary quantized fermionic field
$\chi_{+\sigma_{_B}}(x)$ with an electromagnetic field reads
\begin{eqnarray}
V= \chi^{\dagger}_{+\sigma_{_B} } \left[-c \vec \sigma_{BA}
\cdot{e\over c}\vec A -
M_{BA}(
0)- \sum_i \left. {d M_{BA}\over d p_i'} \right|_{p_i'=0}\right.
\nonumber \\
\times \left( p_i ^{AB} -{e\over c} A_i \right)  - {1\over 2}
\sum_{i,j} \left. {d^2 M_{BA} \over d   p_i' d   p_j'}
\right|_{p_i', \ p_j' =0}
\left.\left( p_i ^{AB} -{e\over c} A_i \right)\, \left( p_j
^{AB} -{e\over c} A_j \right) + \ldots\right]\chi_{+\sigma_{_B} }.
\label{interaction_V_graphene}
\end{eqnarray}
To perform quantum-statistical averaging for the case of non-zero
temperature, we  use a quantum field method developed in
\cite{Varlamov,myConductivity}. After tedious but simple algebra
one can find   the conductivity in  our model:
\begin{eqnarray}
\sigma_{ii}^{O}(\omega
, \ k) =  {\imath e^2\bar{\beta}^2\over  (2\pi c)^2}   \mbox{Tr}\
 \int  \left(1-
  M_{BA}(\vec p )    {\partial^2 M_{BA}    
   \over \partial   p_i ^2}
  \right)
\left(  M \vec v^i(p)
 \, , \,
N\vec v^i(p)
 \right)\,   d\vec {p}   ,
 \label{conduction3}\\
 \sigma_{ll}^{Zb}(\omega
 ,  k)={ \imath e^2\bar{\beta}^2  \over   (2\pi c)^2}
\mbox{Tr }\ \int {
M_{BA}(\vec p)\over 2}  \sum_{i=1}^2 {\partial^2
M_{BA} \over \partial   p_i^{2}}
 \left(M \vec v^i(p) \, , \, N\vec v^i(p) \right) \,   d\vec {p}   ,
 \label{Zitterbewegung_conduction1}\\
 \sigma_{12(21)}^{so}(\omega
 ,  k )= { (-1)^{1(2)} \imath\over 2 }{\imath e^2\bar{\beta }^2\over   (2\pi c)^2} 
 \mbox{Tr }\
 \int
 {
 M_{BA}(\vec p)
 }
{\partial^2 M_{BA}
\over \partial p_{1} \partial p_{2}}
\left( M\vec v\, ^{1(2)}(p)
\, , \, N \vec v\, ^{1(2)}(p) \right)  \sigma_z   d \vec {p}
 \label{spin-orbit-conduction1}
\end{eqnarray}
for the currents $j_i^{O},\ j_i^{Zb},\ j_i^{so}$ respectively. Here
the matrices $M,\ N$ are given by the following expressions:
\begin{eqnarray}
M =  {f[\bar {\beta} ((H(p^+)-\mu)/\hbar)]
- f[\bar {\beta} (H^\dagger (-p^-)-\mu/\hbar)]\over \bar {\beta} 
z^{
}
- \bar {\beta} (H(p^+)/\hbar) + \bar {\beta} (H^\dagger
(-p^-)/\hbar) },\
 N= {\delta\left(\hbar \omega
 + \mu
 \right)\over   (\hbar z
+H(p^+) - H^\dagger (-p^-))\bar {\beta} }.
\nonumber 
\end{eqnarray}
Here $f$ is the Fermi--Dirac distribution, $z=\omega +i \epsilon
$, $\vec p^{\pm}=\vec p \pm \vec k$, $\omega$ is a frequency,
$\mu$ is a chemical potential, $\bar {\beta}$ is an inverse
temperature divided by $c$.

\section{Results and discussion}

\subsection{Band structure simulations}

\begin{figure}[hbtp]
\begin{center}
(a) 
\\
\includegraphics[width=12.5cm,height=8.cm]{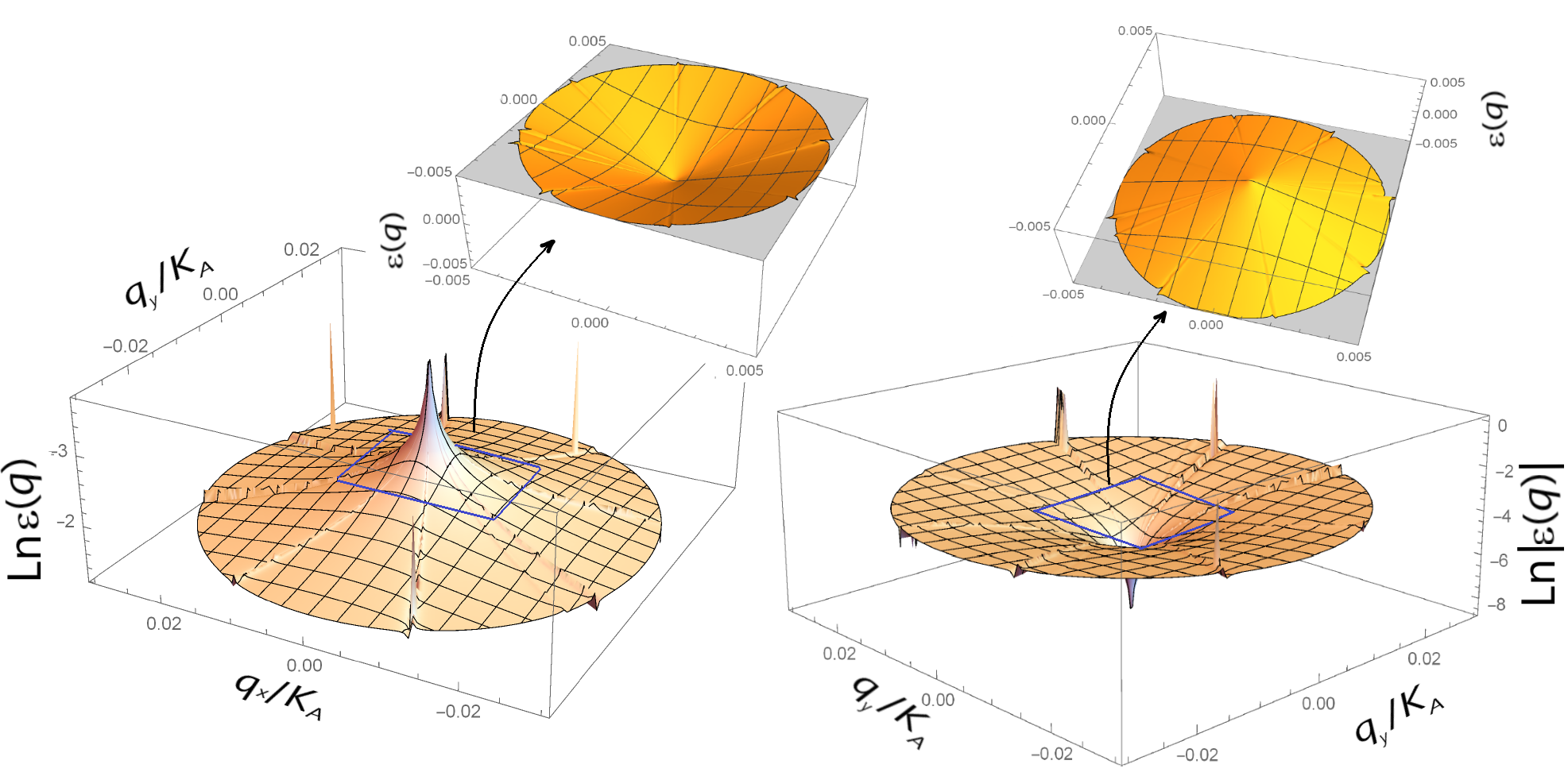}\\
 (b) \\
\includegraphics[width=13.cm,height=6.5cm]{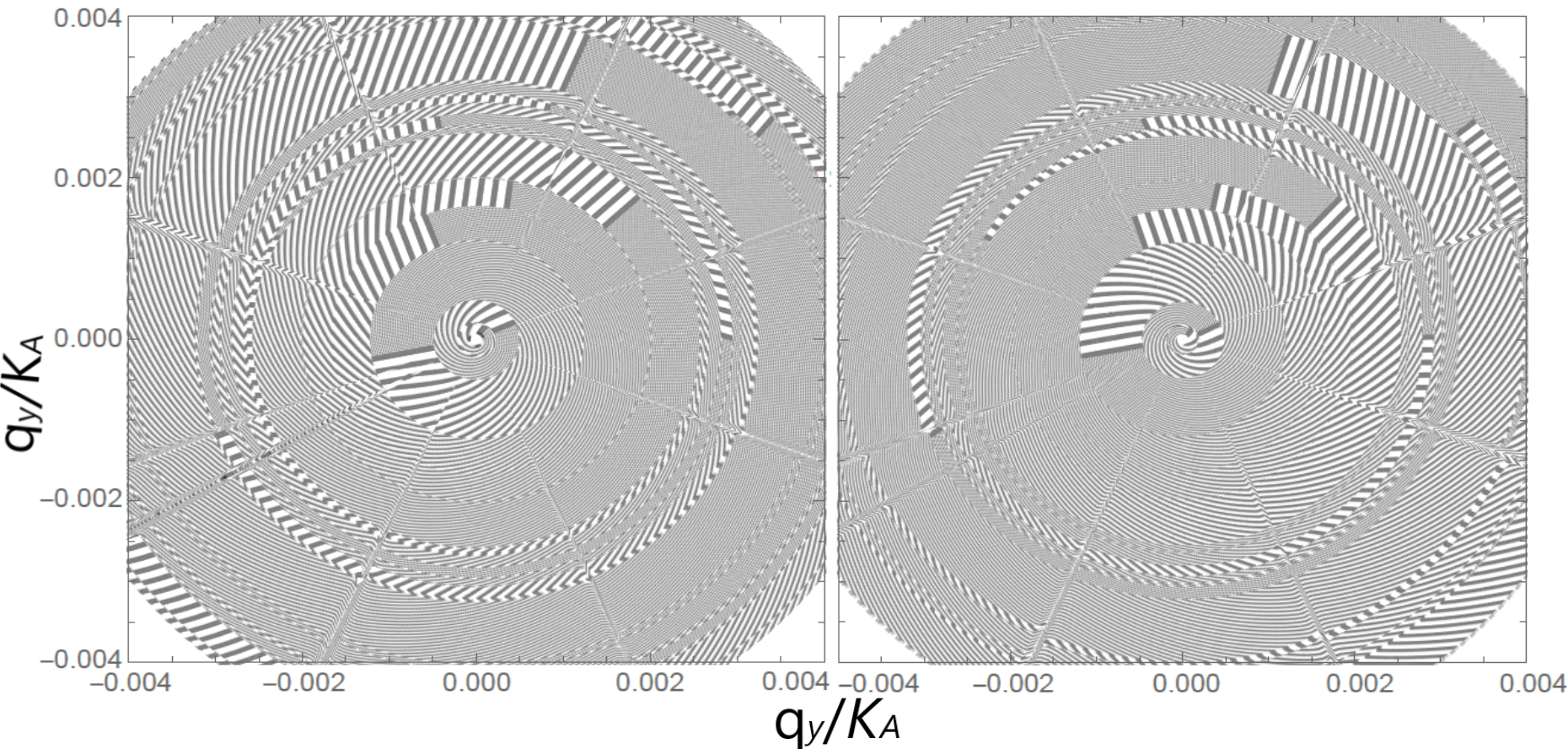}
\end{center}
\caption{
(a) Electron (left) and hole  bands (right) 
 of a quasi-relativistic $N_F=3$-flavors model graphene   with a pseudo-Majorana mass term.
(b)
  The vortex texture in contour plots  of the graphene electron (left)  and hole  (right) bands.}
 \label{band-structure}
\end{figure}

The band structure of graphene within the quasi-relativistic $N_F=3$-flavors model
   has been calculated with the Majorana-like mass term and
is presented in Fig.~\ref{band-structure}a. The  graphene bands are
conical near the Dirac point at   $q(q')\to 0$,  $q=\left|\vec p -
\vec K\right|$ ($q'=\left|\vec p\;\! ' - \vec K'\right|$) where
$\vec p (\vec p\;\! ')$ is a momentum of electron (hole). But,
they flatten at large $q(q')$.

The band structure of graphene within the quasirelativistic  model with
pseudo Majorana charge carriers hosts vortex and antivortex whose
cores are in the graphene valleys $\vec K$ and $\vec K'$ of the
Brillouin zone, respectively (see Fig.~\ref{band-structure}b).
Touching in the Dirac point $K(K')$
the cone-shaped valence and conduction bands of graphene flatten
at large momenta $p$ of the graphene charge carriers
\cite{our-symmetry2020}. It signifies that the Fermi velocity
$v_F$ diminishes drastically to very small values at large $p$. Since eight sub-replicas
of the graphene band near the Dirac point degenerate into the
eight-fold conic band (see Fig.~\ref{band-structure})   the
 pseudo Majorana fermions forming the eightfold degenerate vortex are confined by
  the hexagonal symmetry.
In the state of confinement, the pseudo Majorana fermions
 are linked with the formation of electron--hole pairs
 under the action of the hexagonal symmetry.

\begin{figure}[htbp]
\includegraphics[width=7.0cm]{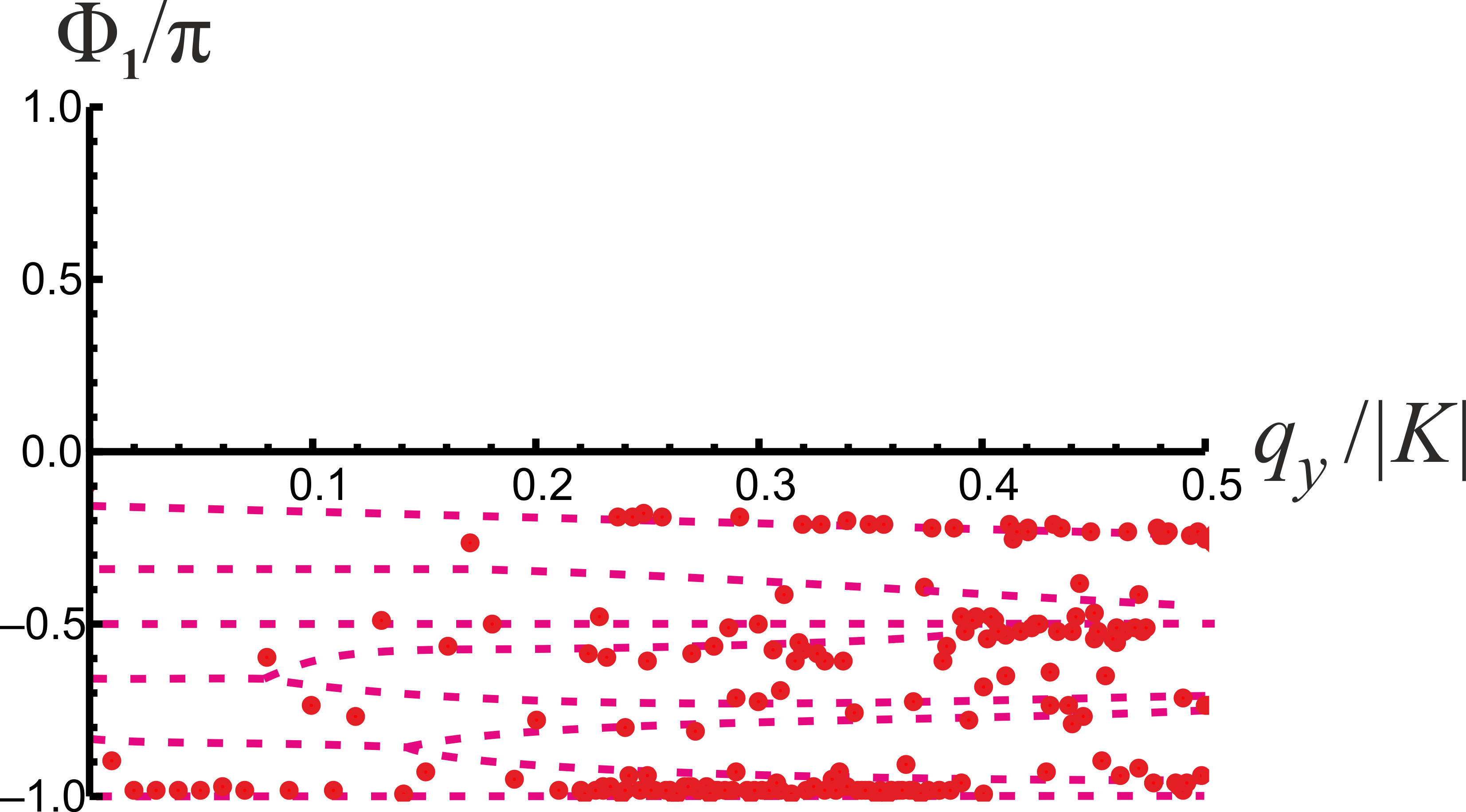} \hspace*{-1.2cm}
\includegraphics[width=7.0cm]{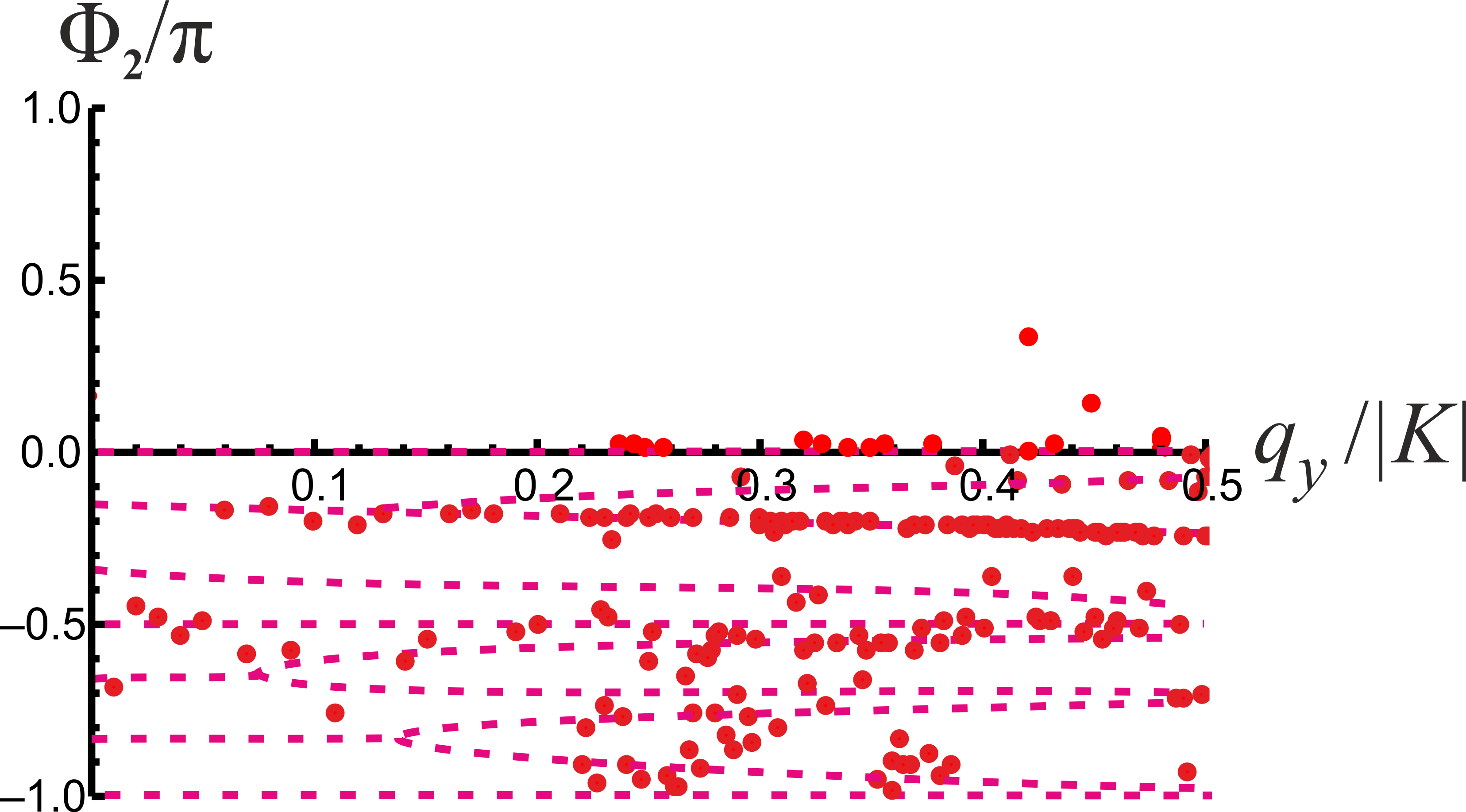} \hspace*{-1.0cm}
\\
\includegraphics[width=7.0cm]{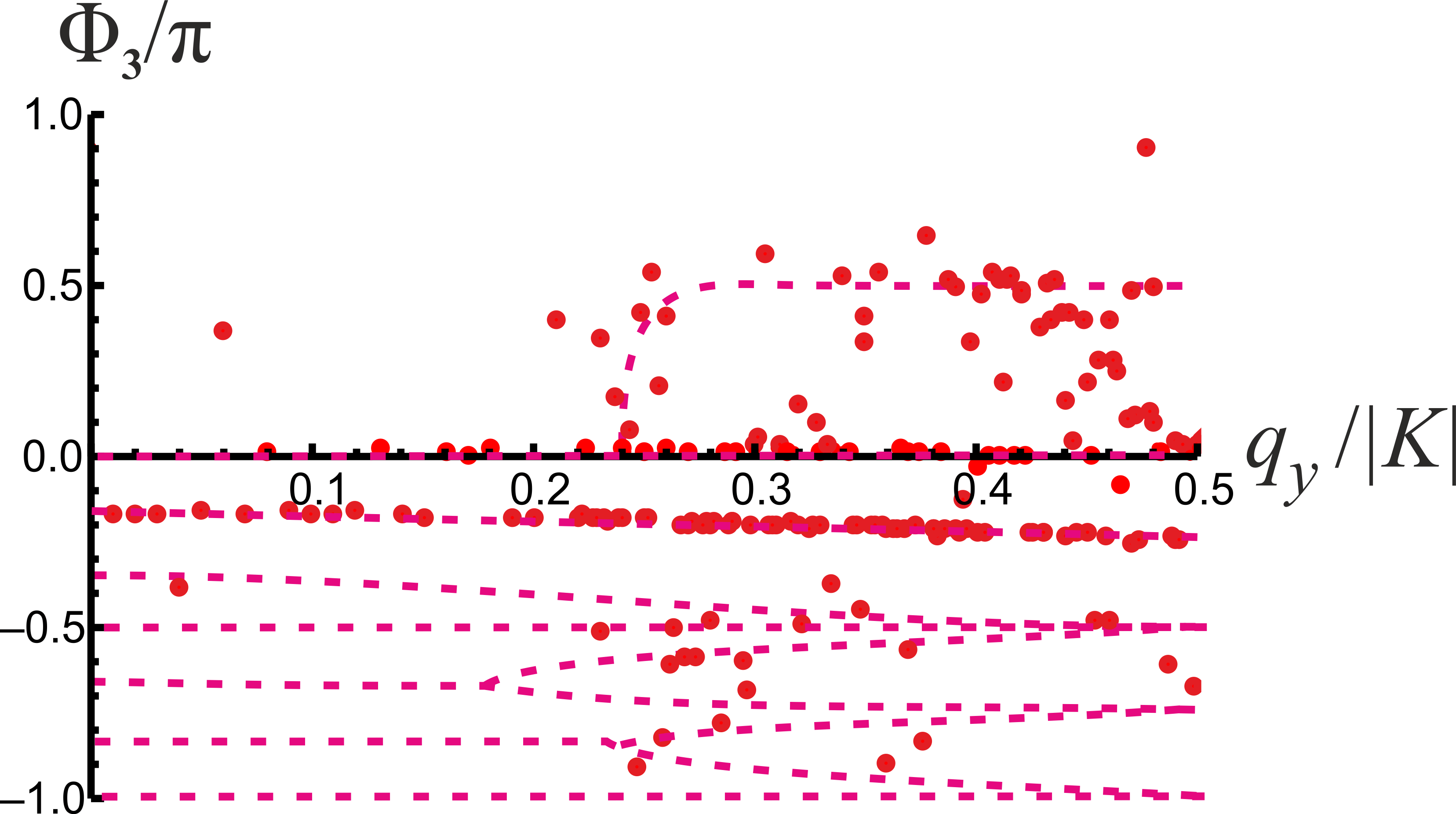}\hspace*{-1.1cm}
\includegraphics[width=7.0cm]{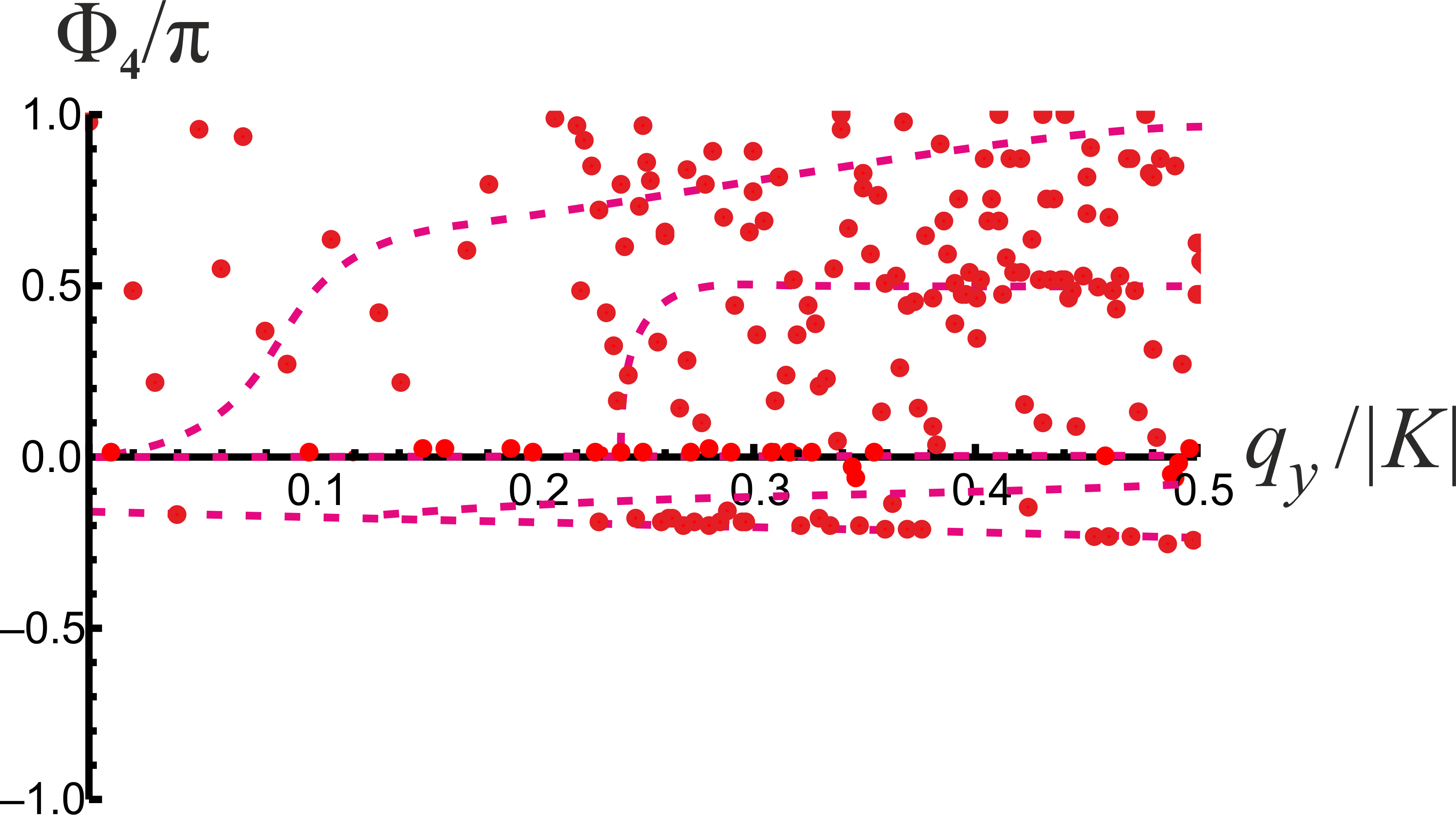}
\caption{
Non-Abelian phases $\Phi_1,\ldots,\Phi_4$ of the
Wilson-loop eigenvalues in the units of $\pi$ at non-zero gauge fields.}
\label{fig-phase}
\end{figure}

The non-Abelian Zak phases of the pseudo Majorana graphene  charge carriers are
nonzero \cite{our-symmetry2020} as it is shown in Fig.~\ref{fig-phase}.
The charge carriers, whose non-Abelian Zak phases are multiples of
$\pi/6$ and constitute the cyclic groups $\mathbb{Z}_{12}$, are
confined near the Dirac point. The $\pi/6$ rotation is equivalent
to a $\pi/2$ rotation due to the hexagonal symmetry of graphene and,
correspondingly, the electron and hole configurations in the
momentum space are orthogonal to each other. It testifies the
metallicity of  zigzag edges and/or zigzag configurations  and
semi-conductivity of armchair edges and/or armchair configurations
transversal to the zigzag configuration in the graphene plane.
 All $\pi (\mbox{p}_z)$-electrons are precessed (move from one valley
to another one) in a same way near the Dirac point because the hexagonal
symmetry levels the transitions between the levels
with different projections $j = \pm 3/2,\pm 1/2$ of the $\pi
(\mbox{p}_z)$-electron orbital momentum $J_{\mbox{p}_z}$
 due to smallness of a
spin-orbital coupling  at momenta $q(q')\to 0$, $\vec q=\vec p-
\vec K $ ($ \vec q\;\!'=\vec p\;\!'-\vec K' $).

The strong spin-orbital coupling  at large momenta $q(q')$ violates the
hexagonal symmetry lifting the degeneration over the projections $j$.
The precessing
of the $\pi (\mbox{p}_z)$-electron proliferates vortices
(antivortices) in the T-shaped configuration of four topological
vortex defects (four antivortices) \cite{our-symmetry2020}.
An atomic chain with two topological defects at the ends
implements a pseudo Majorana particle
\cite{Physics-Uspekhi44-2001Kitaev,JPhysB40-2007Semenoff}. Such
T-shaped configuration of four topological vortex defects (four
antivortices) is three
pseudo Majorana quasiparticles differing in the combinations of
vortex subreplicas that form them.
The number of the pseudo Majorana modes coincides with the number
of  the gauge degrees of freedom ($ N_F = 3 $)  and,
accordingly, all three Majorana modes differ in the flavor.
It signifies that the pair of vortical and antivortical
subreplicas holds one of three flavors.

A feature of the pseudo-Majorana mass term is its vanishing in the valleys $K(K)$.
Outside the valleys one of  two eigenvalues of the pseudo-Majorana mass term entering
the Hamiltonian of the pseudo Majorana fermion turns out to be
practically zero \cite{Grush-Kr2017}.
Therefore, one of the flavored Majorana particles is formed by two
chiral vortex defects, the second one -- by two nonchiral
vortices, and only one of the two vortices is chiral for the third Majorana mode.
Since the flavor is associated with a degree of chirality, let us
call the pseudo Majorana flavor modes as the chiral, semichiral, and
non-chiral pseudo Majorana quasiparticles $V_{ch},V_{sc},V_{nc}$.

The system of Eqs.~(\ref{Majorana-bispinor01},
\ref{Majorana-bispinor02}) for the stationary case can be
approximated by
a Dirac-like equation with the ``Majorana-force`` correction in
the following way.
The operator $\Sigma_{AB}^{-1}$ in (\ref{Majorana-bispinor01})
plays a role of Fermi velocity also: $\hat v'_F=\Sigma_{AB}$. Then
one can assume that there is the following expansion up to a
normalization constant $\left<0\right|\hat v_F \left|0
\right>=\left<0\right|\hat v'_F \left|0 \right>$:
\begin{eqnarray}
\left| \psi_{AB}\right\rangle =\frac{\Sigma_{AB}\left|
\psi^*_{BA}\right\rangle}{\left<0\right|\hat v'_F \left|0
\right>}=
\frac{\Sigma_{AB}\Sigma_{BA}}{ \left<0\right|\hat v'_F \left|0
\right>} \left|
\psi_{AB}\right\rangle 
=\frac{\left\{ \Sigma_{BA}  + \left[\Sigma_{AB},
\Sigma_{BA}\right]\right\} \left|\psi_{AB}\right\rangle}{
\left<0\right|\hat v_F \left|0
\right>} %
    \nonumber \\
\approx \left\{ 1  + {\left(\Delta \Sigma + \left[\Sigma_{AB},
\Sigma_{BA}\right]\right)
\over \left<0\right|\hat v_F \left|0 \right>}\right\} \left| \psi_{AB}\right\rangle 
 \label{permutations1}
\end{eqnarray}
where $\left[\cdot, \cdot \right]$ denotes the commutator, $\Delta
\Sigma = \Sigma_{BA} - \Sigma_{AB} $.
Substituting (\ref{MwFun}, \ref{permutations1}) into the right-hand side of the equation 
\eqref{Majorana-bispinor02}, one gets the Dirac-like equation with
a ``Majorana-force'' correction of an order of quantum-exchanges
 difference  for two graphene sublattices:
\begin{eqnarray}
\left[ \vec \sigma_{2D}^{AB}\cdot \vec p_{BA}-c^{-1} M_{BA} 
\right]
\left| \psi^*_{BA}\right\rangle 
 = \tilde E\left\{ 1 +{\left(\Delta
\Sigma + \left[\Sigma_{AB}, \Sigma_{BA}\right]\right)\over
\left<0\right|\hat v_F \left|0 \right>} \right\}\left|
\psi^*_{BA}\right \rangle \label{Majorana-bispinor1}
\end{eqnarray}
where $\tilde E=E/\left<0\right|\hat v_F \left|0 \right>$.
Now, neglecting the  mass term, we can find the solution of
the equation \eqref{Majorana-bispinor1} by the successive
approximation technique as:
\begin{eqnarray}
\vec \sigma_{2D}^{BA}(\Delta_{\pm, i})\cdot \vec p_{AB}
(\Delta_{\pm, i})\left| \psi_{AB}\right\rangle +{E^{(0)}
\left(\Delta \Sigma (\Delta_{\pm, i})+ \left[\Sigma_{AB}
(\Delta_{\pm, i}), \Sigma_{BA}(\Delta_{\pm, i})\right]\right)\over
\left<0|\hat v_F|0\right>^2 }\left| \psi_{AB}\right \rangle \nonumber \\
=
{E^{(1)} \over \hat v_F}\left| \psi_{AB}\right \rangle . %
\label{variational-Majorana-bispinor}
\end{eqnarray}
It follows from Eq.~(\ref{variational-Majorana-bispinor}), that
the quadratic correction describes a deconfinement of the
pseudo Majorana fermions by SOC  because the ``Majorana-force`` is significant
in the flat regions of the graphene bands where the Fermi velocity trends to zero.

 Thus, the pseudo Majorana fermion graphene model is a topological semimetal.
Resulting in   8 subreplicas  of the graphene band,  the
spin-orbital coupling is  capable to compete with the hexagonal
symmetry at high energies in the flat bands only. The pseudocubic
symmetry  signifies that the electron-hole symmetry of each
graphene band  is separately broken and, correspondingly, the
associated vortex and antivortex Majorana fermions forming
electrons and holes are released by strong SOC.
These free Majorana particles exist in a very narrow energy range
because they reside in the flat regions of the graphene bands.
Since the Fermi velocity
$v_F$ of the free Majorana configurations trends
 to zero, the pseudo Majorana fermions are very heavy ones.

\subsection{Low-frequency dielectric permittivity of graphene}

Let us  investigate  longitudinal conductivity  for low
frequencies  $\omega\to 0$ and  non-vanishing  wave  vectors $\vec
k=\vec p/\hbar - \vec K_{A(B)}$.
The longitudinal   conductivity
$\sigma_L(\omega,{\vec k})$ is determined
 through the conductivity tensor splitting into
longitudinal and transversal terms as \cite{Kraft-Ropke}
\begin{equation}
 \sigma_{ij}(\omega,{\vec k})=
 \left(\delta_{ij}-\frac{k_i k_j}{{\vec k}^2} \right)\sigma_{T}(\omega,{\vec k})+
\frac{k_i k_j}{{\vec k}^2}\sigma_{L}(\omega,{\vec k}); i,j=x,y .
\end{equation}
When choosing $\vec k= (k_x,0)$ or $\vec k= (0,k_y)$, $k_x=k_y=k$ one always has
$$
 \sigma_{xx}(\omega,{\vec k})= \sigma_{L} (\omega,k),\  \mbox{or}  \
 \sigma_{yy}(\omega,{\vec k})= \sigma_{L} (\omega,k).
$$

\begin{figure}[htbp]
(a) \hspace{8cm} (b) \\
\includegraphics[width=7cm,height=4cm]{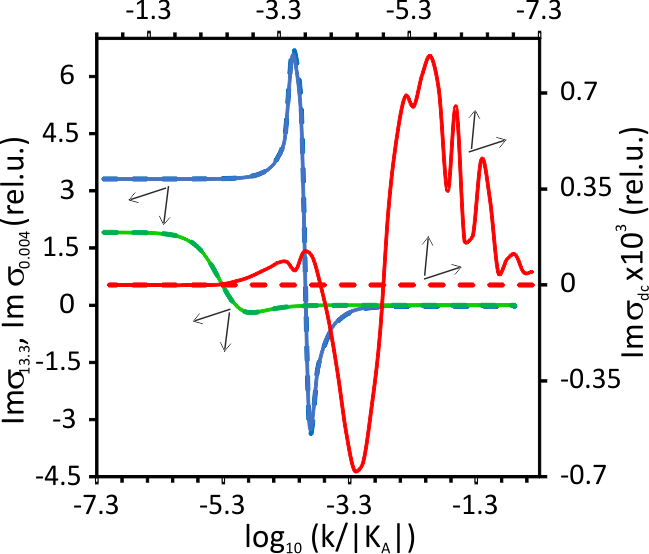} 
\includegraphics[width=6cm,height=3cm]{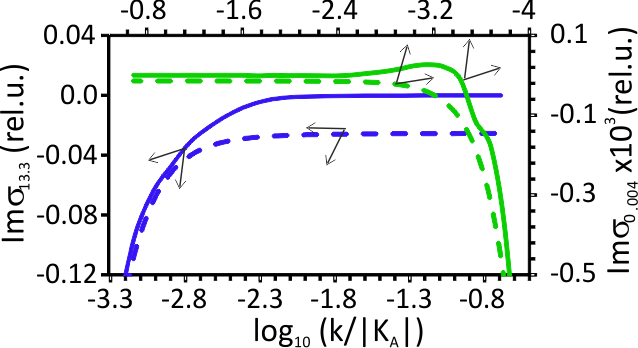}\\
(c) \hspace{8cm} (d) \\
\includegraphics[width=7cm,height=4cm]{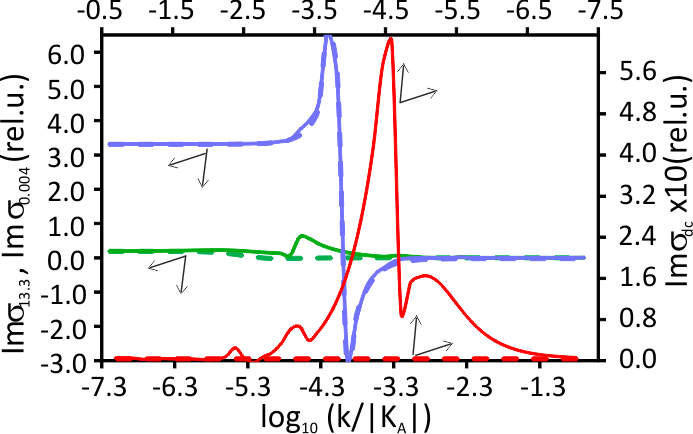} 
\includegraphics[width=6cm,height=4cm]{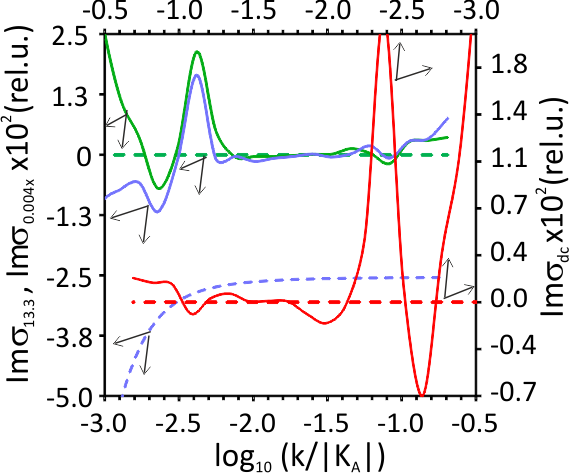}
\caption{Imaginary part $\sigma_{L}^{O}(\omega, k)$ of longitudinal ohmic contribution
to  conductivity vs  wave number $k$, $\vec k = \vec p - \vec K_{A(B)}$ for our pseudo Majorana fermion
$N_F = 3$-model \eqref{Majorana-bispinor01},\eqref{Majorana-bispinor02}
without  (a,b) and with (c,d) the pseudo-Majorana mass term
(solid curves) and for the massless Dirac fermion model
 \cite{Varlamov} (dashed curves),
at the temperature 100~K and the frequencies $\omega$: $13.3$~K
($0.27 $~THz, blue color), $0.004$~K ($83$~MHz, green
color), $10^{-10}$~K ($2.08$~Hz, red color); the chemical
potential equals to 1~K. The inset (b) to the figure (a)
demonstrates the $\Im m\ \sigma_{L}^{O}$ in a neighborhood of
large $k$ at the frequencies 13.3 and $0.004$~K; the
inset (d) to the figure (c) demonstrates the $\Im m\
\sigma_{L}^{O}$ in the region of large  $k$ at the all frequencies
for the case taking into account of the pseudo-Majorana mass term.
$\sigma_L^O (\omega, k) $ is measured in the relative unit $e^2/h$
and labeled as $\Im m \ \sigma_{13.3},\Im m \ \sigma_{0.004}$ and $\Im m \ \sigma_{dc}$
for the frequencies 13.3, $0.004$, and $10^{-10}$~K, respectively. } \label{dc-conductivity}
\end{figure}

Now let us calculate the low-frequency dielectric permittivity  $\epsilon(\omega, \vec r)$. To do it one
has  to perform the inverse Fourier transformation
\begin{eqnarray}
\sigma (\omega, \vec r)={1\over (2\pi)^2} \int  {\sigma^O
_{L}}(\omega ,  k) e^{i \vec k\cdot \vec r} d^2k
\label{dc-complex-conductivity-estimation0}
\end{eqnarray}
and then to substitute the transformation into the following expression:
\begin{eqnarray}
\epsilon (\omega, \vec r)=\Re e \left[1-4\pi \imath {\sigma (\omega , \vec r)\over \omega}  \right] =
1+{4\pi \over \omega} \Im m\ \sigma (\omega , \vec r) .
\label{dc-permoittivity-estimation-0}
\end{eqnarray}
Here $\omega$ is a cyclic frequency.
We consider the effects of  spatial dispersion on the imaginary part $\Im  m \ \sigma ^O_L(\omega,k)$
of the longitudinal complex conductivity at the low frequencies:
$\omega=10^{-10}$, $0.004$, 13.3~K~(kelvin) for the
massless pseudo Dirac graphene fermion model with the number of flavors
$N_F=2$  (pseudospin and spirality) and our  graphene model with the
$N_F=3$ flavors. Conductivity for frequencies in  Hertz range, for
example, 2.08~Hz ($ \omega = 10^{-10}$~K) can be
considered as a conductance for direct current.
The numerical results are presented in Fig.~\ref{dc-conductivity} and Table~1.

\begin{table}[htbp]
\small\caption{Asymptotic  behavior of the
longitudinal-conductivity imaginary part  $\Im m\ \sigma_L^O (\omega, k) $ in the massless
pseudo Dirac  fermion  monolayer-graphene $N_F=2$-model  and  in the
pseudo Majorana fermion monolayer-graphene  $N_F=3$-model for the case taking into account of
the pseudo-Majorana mass term
(mass case) and without the mass term (massless)  at different frequencies $\omega$
and the large and small wave numbers $k_\infty $ and  $k_0$,
 namely, at  $0.2 |K_A|$ and $6.5\cdot 10^{-8}|K_A|$, respectively; at
the temperature 100~K and the chemical potential 1~K;
$\sigma_L^O (\omega, k) $ is measured in the relative unit $e^2/h$. }
\begin{tabular}{|p{0.5in}|p{0.6in}|p{0.6in}|p{0.6in}|p{0.6in}|p{0.6in}|p{0.6in}
}
\hline \hline
 {\multirow{3}{*}{$\omega$, K}}
 & \multicolumn{3}{|p{1.8in}|}{$\Im m$ $\sigma_L^O(\omega, k_\infty)$}  & \multicolumn{3}{|p{1.8in}|}
 {$\Im m$ $\sigma_L^O(\omega, k_0)$} 
\\
\cline{2-7
} & \textbf{N$_F$=2} &
\multicolumn{2}{|p{1.2in}|}{\hspace{1cm} \textbf{N$_F$=3}} &
\textbf{N$_F$=2} &
\multicolumn{2}{|p{1.2in}|}{\hspace{1cm}\textbf{N$_F$=3}}
\\
\cline{2-7
}  &  & massless & mass case &  & massless & mass case 
\\
\hline \hline 13.3 & -0.025 & $3 \cdot 10^{-5}$ & $7.5\cdot
10^{-3}$ &
3.312 & 3.308 & 3.32 
\\
\cline{1-7} 
$0.004$ & -$1.4\cdot 10^{-5}$ &
$5.4\cdot
10^{-8}$ & $3.6\cdot 10^{-3}$ & 0.191 & 0.191 & 0.194 
\\
\cline{1-7} 
$10^{-10}$ & 0 & $6.3\cdot 10^{-8}$ &
$1.8\cdot 10^{-3}$ &
        -$1.4\cdot 10^{-8}$ & $4.8\cdot 10^{-5}$ & $2.5\cdot 10^{-3}$
        \\
\hline \hline
\end{tabular}
\end{table}

The function $\Im  m \ \sigma ^O_{L}(\omega,  k)$   for the massless
pseudo Dirac  fermion graphene model is a constant function $\left. \Im  m \
\sigma ^O_{L}(k)\right|_{k\to \infty }\equiv \Im  m \ \sigma
_{k_{\infty}}$ at large wave numbers. The  $ \Im m \ \sigma _{k_{\infty}}$
is negative constant ($\Im m\ \sigma _{k_{\infty}}<0$) for the frequencies $\omega=0.004$, 13.3~K
  and a zero-values function ($\Im  m \ \sigma _{k_{\infty}}=0$) at $\omega=10^{-10}$~K
  (see Table~1 and Fig.~\ref{dc-conductivity}).
But, the imaginary part of the longitudinal complex conductivity in the $N_F=2$-model  becomes a  positive constant
 $\left. \Im  m \ \sigma ^O_{L}(k)\right|_{k\to 0}\equiv \Im  m \ \sigma _{k_{0}}>0$
at small values of $k$, $k\ll 1$ for the frequencies $\omega=0.004$, 13.3~K.
As the Table~1 and Fig.~\ref{dc-conductivity} show the function
$\Im  m \ \sigma ^O_{L}(\omega,k)$ in  the $N_F=2$-model
changes its sign in a very narrow range of $k/K_A$. 
Since $\Im  m \ \sigma ^O_{L}(\omega,k)$
 is  constant almost everywhere for the $N_F=2$-model of graphene, does not oscillate,  and
$e^{i \vec k \cdot \vec r}$ enters  the integrand of the expression
\eqref{dc-complex-conductivity-estimation0} the $\Im m \sigma (\omega, \vec r)$ is equal to zero and,
correspondingly,
 the $ \epsilon (\omega, \vec r) $ is  equal to 1 in this graphene model.
It signifies that the graphene  with  the massless pseudo Dirac charge carriers is not polarized and
plasmon oscillations in the $N_F=2$ model are not observed. But this prediction for the dc case contradicts the experimental facts that value of
the graphene dielectric constant $\epsilon_G$
is in the range 2--5.

The function $\Im  m \ \sigma ^O_{L}(\omega,k)$   for the pseudo Majorana graphene $N_F=3$-models both with
and without the Majorana mass term  trends to non-negative values at $ 
k\to \infty$. It signifies that the polarization states can emerge in the
graphene $N_F=3$-models. The function $\Im  m \ \sigma^O_{L}(\omega,k)$  for the  pseudo Majorana fermion
graphene $N_F=3$-model
without  the pseudo-Majorana mass term  is constant 
at large wave numbers for $\omega=$13.3~K only (see Figs.~\ref{dc-conductivity}a,b).
The $\Im  m \ \sigma ^O_{L}(\omega,k)$  is  weakly or
strongly oscillating for $\omega=0.004$ and $10^{-10}$~K, respectively.
Since $\Im  m \ \sigma^O_{L}(\omega,k)$ for $\omega=0.004$ and 13.3~K
 is practically  constant except for a very narrow interval, then as well as
for the $N_F=2$ model, the $\epsilon(\omega, \vec r)$ is equal to 1
and, correspondingly, the graphene  with  the chiral pseudo Majorana charge carriers is
not dielectrically polarized at these frequencies.

Let us examine the $N_F=3$ model without the pseudo-Majorana mass term
in the dc case of $\omega= 10^{-10} $~K. In this case, since the
$\Im m \ \sigma ^O_{L}(\omega,k)$ has extrema and, oscillating, tends
to small positive values, it behaves like a linear combination of
functions
 $\Im m\ \sigma _{k_{max_1}}{\sin (k-k_{max_1})\over (k-k_{max_1})}$ and
$\Im m\ \sigma _{k_{max_2}}{\sin (k-k_{max_2})\over (k-k_{max_2})}$.
Such sort of functions can be considered as a finite approximation of
the Dirac $\delta-$function, and the coefficients $\Im m\ \sigma _{k_i},\ i=1,2$ are called
intensity or spectral power of the $\delta-$functions.
Then, the dc dielectric
permittivity $\epsilon^{dc}(\vec r)\equiv \left.\epsilon(\omega, \vec r)\right|_{\omega =10^{-10}} $
 for the $N_F=3$ model with the
chiral pseudo Majorana fermions can be approximated by the following expression:
\begin{eqnarray}
\epsilon^{dc}(\vec r)  \approx 1+ {4\pi \over \omega }{1\over 2\pi} \int \left[
\Im m\ \sigma _{k_{max_1}} \delta (k-k_{max_1}) -  \Im m\ \sigma _{k_{max_2}} \delta (k-k_{max_2})
 \right] e^{i \vec k\cdot \vec r} d k.
 \label{dc-permittivity-estimation-zero-mass}
\end{eqnarray}
Since, according to the simulation shown in Fig.~\ref{dc-conductivity}a,
$\Im m\ \sigma _{k_{max_1}}, \Im m\ \sigma _{k_{max_2}}$
differ slightly from each other, the $ \epsilon ^ {dc} (\vec {r}) $ gains a value close to 1.

Now let us examine the $N_F=3$ model with the non-zero pseudo-Majorana mass
term. In this case, the $\Im  m \ \sigma ^O_{L}(\omega,k)$ strongly oscillates for the all frequencies.
In the dc limit ($\omega=10^{-10} $~K) the $\Im  m \ \sigma ^O_{L}(\omega,k)$ possesses
one maximum at $k_{max}/K_A\approx 10^{-3.3}$ only and trends to approximately to the same value
at $k\to 0, \infty $ (see Figs.~\ref{dc-conductivity}c,d and Table~1). Correspondingly, the dc dielectric
permittivity $\epsilon^{dc}({\vec r})$ for the $N_F=3$ model with the chiral anomaly may be approximated
by the integral with only one Dirac $\delta$-function entering the integrand:
\begin{eqnarray}
\epsilon^{dc}(\vec r) \approx 1 + {2\over \omega } \int  \Im m \ \sigma
_{k_{\infty}} \delta (k-k_{max}) e^{i \vec k\cdot \vec r} d k.
 \label{dc-permittivity-estimation-nonzero-mass}
 \end{eqnarray}
Since $k_{max}\neq 0$ the dc dielectric permittivity
$\epsilon^{dc}(\vec r)$ is a periodic function with amplitude $
\sim 1.7 $ that exactly coincides with the graphene dielectric
constant $ \epsilon_G $ at the small charge density $ n \to 0 $.
The $\Im  m \ \sigma ^O_{L}(\omega,k)$ at  $\omega=0.004$~K
behaves similarly to $\Im  m \ \sigma ^O_{L}(\omega,k)$ at
$2.08$~Hz. Since the  $\epsilon (\vec r)$ is periodic, it can take
on  values close to zero that is a signature of a plasmon resonance
in graphene.

The $\Im  m \ \sigma ^O_{L}(\omega,k)$ at  $\omega=13.3 $~K has two extrema and trends
to the different values at $k\to 0, \infty $ (see Fig.~\ref{dc-conductivity}c and Table~1).
 It means that the low-frequency dielectric permittivities $\left.\epsilon(\omega,\vec r)\right|_{\omega
=13.3}\equiv \epsilon_{13.3}(\vec r)$ for the $N_F=3$ model with the chiral anomaly may be approximated
by the integral with the sum of the  difference between two different Dirac $\delta$-functions and
a Heaviside $\Theta$-function entering  the integrand as
\begin{eqnarray}
\epsilon_{13.3}(\vec r)  \approx 1+ {4\pi \over \omega }{1\over 2\pi}
\int e^{i \vec k\cdot \vec r}\nonumber \\
\times \left[ {\Im m\ \sigma
_{k_{max_1}}\over 2} (\delta (k-k_{max_1}) + \Theta (k_{max_1}-k))
-  \Im m\ \sigma _{k_{max_2}} \delta (k-k_{max_2})
 \right]  d k.
 \label{dc-permittivity-estimation-nonzero-mass}
\end{eqnarray}
According to the simulation results, the $\epsilon_{13.3}(\vec r), \epsilon_{0.004}(\vec r)$, and
$\epsilon ^{dc}(\vec r)$ are the same in the $N_F=3$ model with the chiral anomaly.
Thus,  at the high  frequencies (0.27~THz)  the charge density of the
the $N_F=3$ model with the non-zero pseudo-Majorana mass term is polarized both anomalously and by ordinary
topologically-trivial way.

\section{Conclusion}

So, four vortex and four antivortex defects forming three
pseudo Majorana fermions are confined by the hexagonal symmetry.
Deconfinement of the Majorana modes stems from a competition
between spin-orbital coupling and hexagonal symmetry.
The non-Abelianity of the Zak phase for the Majorana fermions
indicates the existence of  polarized  charge density in graphene.
The dielectric permittivity of the model graphene without
the pseudo-Majorana mass term tends to values near to unity  in the low-frequency limit
because all vortical and antivortical  pseudo Majorana graphene modes  are chiral ones.
The phenomenon of a chiral anomaly is observed for the graphene
charge  carriers with the non-zero pseudo-Majorana mass term since one
 of the vortices in  the pseudo Majorana vortex pair can acquire a nonzero mass.
The breaking of the gauge symmetry leads to the appearance of
nonzero polarization of a graphene region that reveals itself
in  periodic dependence of  the dc dielectric permittivity.
The periodicity of the $\epsilon (\vec r)$ can clarify an emergence of plasmon oscillations
 in graphene.
The dielectric polarization in the graphene $ N_F = 3 $-model occurs due to
the fact that electrons and holes,
bending around topological defects, diverge.
Correspondingly, topological defects prevent the electron-hole annihilation by creating
 an effective polarization vector. Our estimate of the low-frequency graphene permittivity
gives $\epsilon_G\sim 1.7$ and, accordingly, is in excellent agreement with the electrophysical experimental data.




\section*{Conflicts of interest}
{The authors declare no conflict of interest.
}







\end{document}